# Implementing commercial inverse design tools for compact, phase-encoded, plasmonic digital logic devices

Michael Efseaff,[a] Kyle Wynne,[a] Krishna Narayan,[a] Mark C. Harrison[a, *]

[a]Chapman University, Fowler School of Engineering, 1 University Dr., Orange, USA, 92866

**Abstract**. Numerical simulations have become an essential design tool in the field of photonics, especially for nanophotonics. In particular, 3D finite-difference-time-domain (FDTD) simulations are popular for their powerful design capabilities. Increasingly, researchers are developing or using inverse design tools to improve device footprints and performance. These tools often make use of 3D FDTD simulations and the adjoint optimization method. In this work, we implement a commercial inverse design tool with these features for several plasmonic devices that push the boundaries of the tool. We design a logic gate with complex design requirements, as well as a y-splitter and waveguide crossing. With minimal code changes, we implement a design that incorporates phase-encoded inputs in a dielectric-loaded surface plasmon polariton waveguide. The complexity of the requirements in conjunction with limitations in the inverse design tool force us to make concessions regarding the density of encoding and to use on-off keying to encode the outputs. We compare the performance of the inverse-designed devices to conventionally-designed devices with the same operational behavior. Finally, we discuss the limitations of the inverse design tools for realizing complex device designs and comment on what is possible at present and where improvements can be made.

*Mark C. Harrison, E-mail: mharrison@chapman.edu

## 1 Introduction

Over the past two decades, using simulation design tools has become the standard approach for new nanophotonic designs. In particular, 3D finite-difference time-domain (FDTD) simulations are very popular. These tools allow for fast iteration and robust exploration of the design space at a low cost[1–5]. Furthermore, they tend to be relatively easy-to-use and accurate, making them accessible to photonics researchers and designers with a wide variety of backgrounds. Therefore, 3D FDTD simulation tools are a valuable first step for design and a strong theoretical complement to experimental results.

More recently, inverse design tools have begun to emerge as a method of finding optimized designs quickly without needing to explore the entire design space[6–11]. These tools allow researchers to define a figure of merit (FOM) and use the tool to find optimized, non-intuitive designs which maximize or minimize that figure. They often implement the adjoint method in conjunction with FDTD or FDFD (finite-difference frequency-domain) simulations to perform optimizations. The FDTD software we use is Lumerical, one of the most popular commercial FDTD software packages that includes a packaged inverse design tool which leverages an integration with Python code[12]. This tool is designed to be easy-to-use and robust and is therefore suitable for researchers who may not be able to write their own inverse-design algorithm, which can be quite difficult to implement. Despite this promise, these commercial inverse design tools have some limitations when it comes to certain complex design considerations.

Inverse design tools often have restrictions on how the FOM can be defined and how many different scenarios can be optimized. They are also often targeted at specific materials or architectures, such as silicon photonics. Additionally, although the underlying code implementing the adjoint optimization may be provided (as is the case with Lumerical's tools), this code is often quite long and complicated. In fact, making modifications to the code could in itself constitute a significant undertaking and may be outside the project scope of designers who simply want to leverage inverse design as a tool. This leaves the question of how useful these tools are for solving complex design problems without modifying the source code. Furthermore, even if they can be used for more complex designs, there is a question of what limitations of the tools will be necessary to overcome.

In the present work, we leveraged Lumerical's inverse design tool to design a compact plasmonic device with logic-gate behavior that acts on phase-encoded inputs. The device is made using dielectric-loaded surface plasmon polariton (DLSPP) waveguides. Both the phase-encoded inputs and the plasmonic nature of the device presented unique challenges for implementing the commercial inverse design tool, which is primarily intended for silicon photonics. We found solutions to implementing the inverse design tool within the tool's normal constraints rather than within the Python code containing the underlying algorithm, modifying only the base simulations and the accompanying setup script responsible for launching the tools. In addition, we compare the performance of the inverse design tool for designing other common photonic circuit components with the same DLSPP waveguide architecture. The inverse-design optimized devices are shown in 3D renderings in Fig 1.

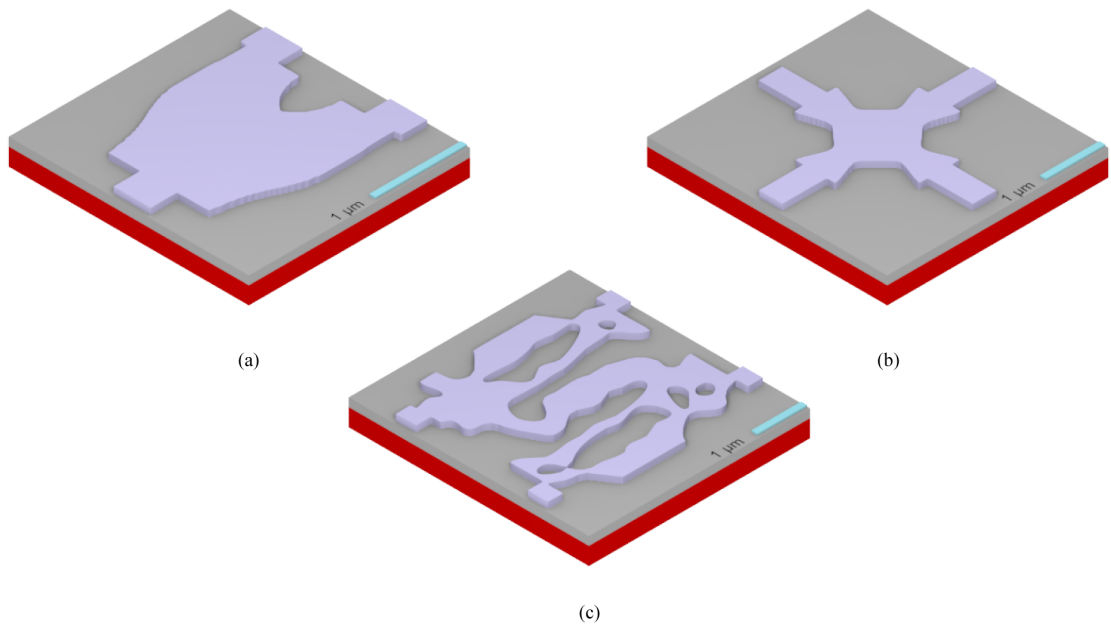

**Fig. 1** 3D renderings of the inverse-designed optimized devices: (a) y-splitter, (b) waveguide crossing, and (c) XOR gate. The devices are not set to the same relative scale.

## 2 Background

There are many simulation tools available (such as Optiwave[13] and OmniSim[14]) as well many inverse design tools (such as Spins[15] and TopOpt[16]), and many researchers create their own inverse-design tools and methods. We considered both simulation method and the packaged inverse design tools before deciding to use Lumerical's inverse design tool, LumOpt. FDFD simulation software is typically restricted to 2D simulations and is better able to handle material dispersion. FDFD methods are useful for small structures that are highly resonant. Finite element method (FEM) simulations are better for time-harmonic problems and are not as useful in simulating propagating light. The FDTD simulation method, used by Lumerical, Optiwave, and OmniSim, is a very generalized method and can easily handle large, 3D devices. Additionally, the use of time when calculating finite-difference equations allows users to observe the fields evolve as a solution is being reached. Based on our desire to create fabricable, non-resonant, 3D devices, inverse design tools using an FDTD solver are preferable.

Inverse design tools tend to operate in a very similar manner to each other, so we evaluated them based on their relative integration with the simulation software. Spins, an inverse design tool created at Stanford, is an open-source tool written in Python integrated with an open-source FDFD solver also developed by the same Stanford group. While Spins allows users to define their own FOM, an advantage compared to LumOpt, being integrated with an FDFD solver limits its applicability to our goal of designing plasmonic waveguide devices. TopOpt is another inverse design tool that is written in MATLAB and useable with COMSOL. While TopOpt is popular and has a number of associated applications, COMSOL uses FEM simulations instead of the FDTD method. COMSOL has a beam propagation method module available as well, but this method can struggle with simulating larger devices and TopOpt is only designed to work with FEM simulations. The integration between COMSOL and MATLAB is also limiting, partially due to the complexity of COMSOL and because MATLAB is a separate piece of commercial software which also must be purchased. Therefore, we chose to use Lumerical FDTD in combination with its inverse design package LumOpt as a general-purpose tool which best suits our application.

We focused on plasmonic devices for a number of reasons. An excellent case has been made for the use of plasmonics to act as active elements in a hybrid circuit with silicon photonics acting as passive waveguides in large-scale photonic integrated circuits (PICs)[17]. Plasmonic devices offer tight confinement of the optical fields and shorter wavelengths than in fully dielectric waveguides, which allows them to have smaller footprints than their silicon photonic counterparts. They also interface easily with electronic circuits, which is desirable for many active components[18]. We selected a DLSPP waveguide architecture (Fig. 2a) for its relatively low propagation loss and ease of fabrication[19]. The architecture consists of a 500 nm layer of silver on a silicon substrate with a patterned dielectric layer on top of that. We used $SiO_2$ for the dielectric layer, but other transparent dielectrics such as optical polymers could be used as well[20,21].

An important application of photonic logic devices is for use in basic processing and routing in fiber optic communication networks[18,22]. Many of these networks take advantage of dense encoding schemes for transmission, such as quadrature amplitude modulation (QAM) or phase-shift keying (PSK). However, dense encoding schemes typically have to be translated to a normal binary encoding when processing information electronically. Designing photonic devices that operate on denser encoding schemes would preclude that requirement, saving power and increasing throughput[23]. For optical processing, PSK encoding schemes are preferable to other, denser options, because PSK has continuous, uniform, and unambiguous transitions between encoded states. Therefore, we choose to design our devices to operate on a binary PSK (BPSK) encoding scheme, where two different phase states of the signal represent digital 0 and digital 1 (Fig. 2b).

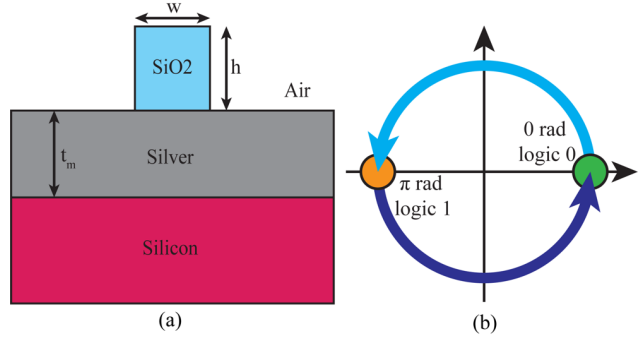

**Fig. 2** (a) Cross section of the single-mode waveguide portion of the devices being simulated, where w = 400 nm, h = 440 nm, and $t_m$ = 500 nm. (b) Binary phase-shift keying diagram. A relative phase of 0 radians corresponds to a logic 0, and π radians corresponds to a logic 1.

The goal of this work was to use inverse design tools to implement digital logic with photonic devices. Although inverse design has been shown for nonlinear devices[8], the Lumerical inverse design tools only work for linear designs[24]. While linear device designs using interference have shown logic function behavior with amplitude-encoded inputs, nonlinear optical processes are required to achieve most logic functions with phase-encoded inputs. An exception to this is the XOR gate[25], which can be demonstrated with linear functions. XOR logic gates are useful for a number of computing applications, and the typical behavior of a two-input XOR gate is that the output is true (1) when the inputs are different from each other, and false (0) otherwise. This behavior, as it was implemented in our devices with PSK encoded inputs and amplitude encoded outputs, is represented in Table 1. Due to the limitations of inverse design and the usefulness of XOR gates for digital logic, we choose to implement this logic function in our design.

**Table 1** XOR Device Truth Table

| A | B | XOR | XNOR |
|---|---|-----|------|
| 0/0 rad | 0/0 rad | 0/OFF | 1/ON |
| 0/0 rad | 1/π rad | 1/ON | 0/OFF |
| 1/π rad | 0/0 rad | 1/ON | 0/OFF |
| 1/π rad | 1/π rad | 0/OFF | 1/ON |

Y-splitters are a commonly used integrated device that take an input signal and split it to two outputs. In a typical design the input signal passing through the y-splitter is split evenly, causing 50% of the signal's power goes to each output, although this is often tuned for different splitting ratios. By using inverse design and a plasmonic structure, we can create a y-splitter with a small footprint, while keeping plasmonic losses manageable. Minimizing loss is important to a y-splitter design as the signals can already only reach a max of 50% power at the outputs, so further loss can result in signal degradation.

Waveguide crossings are another basic photonic device that allows two optical signals to pass through one another at a junction while minimizing loss caused by the disturbing the waveguide modes due to the crossing. By using inverse design, we can reduce the footprint of the waveguide crossing and also increase the transmitted power of both signals to the output waveguides after the crossing junction compared to a conventional design.

## 3 Methods

As previously mentioned, we implemented Lumerical's inverse design tool, a Python suite called LumOpt, to achieve our designs. Specifically, we performed topology optimizations. LumOpt leverages the adjoint method for inverse design optimization[26]. This method allows the gradient of a design space to be calculated with only two simulations: the normal, or forward simulation, and the reverse, or adjoint simulation. In the adjoint simulation, the location and direction of the light source and output monitor are swapped. The design is optimized using a figure of merit (FOM) that matches the light mode somewhere in the device, typically an output port. Optimization of different outputs is also possible by setting up a unique base (input condition) for each FOM.

Inverse design optimizations using LumOpt can take a significant amount of time to complete, even on simulation computers with a large amount of processing power. This is compounded by the fact that we had to use a uniformly spaced mesh for our simulations. The automatically generated non-uniform mesh was smaller than the inverse design topology resolution near the metal interface, so we used a fixed mesh to avoid errors in the gradient calculation step. Unfortunately, this uniform mesh increased the time to complete simulations.

In order to speed up the overall design process for the XOR gate, we first optimized the design using 2D simulations. This allowed us to run each individual simulation faster and therefore have a larger number of simulations run, resulting in a better optimization. However, the 2D designs are not as accurate as 3D designs, since they assume infinite extent in the dimension not simulated. This problem is made worse by the fact that we used plasmonic devices, and the plasmonic confinement was in the dimension not simulated. Therefore, 3D optimization is necessary.

After running the 2D optimization, we used the results as the starting point of the 3D optimization. This allowed the optimization to converge more quickly, requiring fewer rounds of optimization and therefore fewer simulations and less time overall. We also found that the results were better when we optimized with the 2D results as the starting condition as opposed to the default starting condition for optimizations, which is the optimization region having a refractive index halfway between the waveguide index and background index.

The design we optimized had two input waveguides 400 nm wide and spaced 2.5 µm apart. The optimization region was 3.5 µm by 3.5 µm, and we set the radius filter of the inverse design tool to 200 nm. The radius filter is used for optimizing at the end of the design process to ensure that all features have the specified minimum radius which in turn ensures they are manufacturable. The design had two output waveguides that were also spaced 2.5 µm apart and were 400 nm wide.

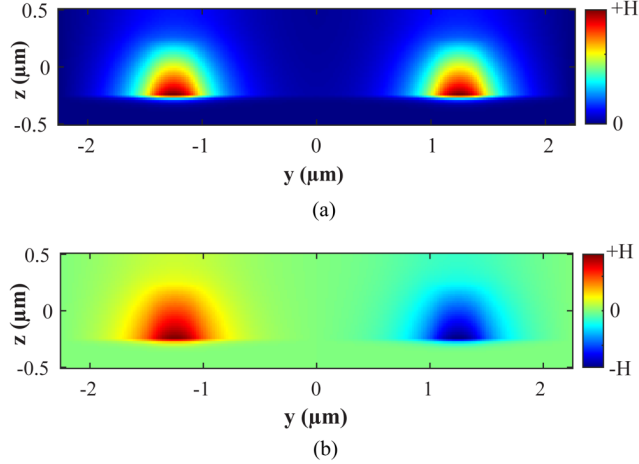

**Fig. 3** In order to simulate different phase relations at the input we used (a) a symmetric coupled mode for in-phase inputs and (b) an anti-symmetric coupled mode for out-of-phase inputs.

The inverse design tool has some important limitations that we had to accommodate in order to realize our goal. First, the adjoint method is implemented in Lumerical such that each base simulation must use only one input light source. Having only one source is an obvious problem for designing a two-input logic gate, which requires two inputs that vary relative to one another. We overcame this limitation by using a single source object in Lumerical and selecting symmetric and anti-symmetric coupled modes for the inputs of different base conditions which are then co-optimized (Fig. 3). These two cases represent inputs that are in-phase with each other and $\pi$ radians out of phase with each other, which are suitable for the two conditions we need to define XOR behavior if we are using a BPSK encoding. In order to validate this approach, we re-ran the simulation with two separate single-mode sources after the optimization was complete to verify it had the same behavior as with the single coupled-mode source used for the optimization. Although we can design with BPSK encoded inputs, we are unable to have higher degrees of phase-shift keying due to this limitation.

Next, we are unable to directly select for phase of the signal at the output. Ideally, we would select for phase conditions so our device uses BPSK encoding at both input and output ports.

Unfortunately, the mode-matching requirement for the FOM means that we can only select for amplitude at the output, as well as mode if we have a multi-mode waveguide. (In our case, the output ports were single mode waveguides). While it would be possible to use the same trick we used for the inputs, with two outputs that have a coupled mode, the result would be a trivial design. Therefore, the outputs of our XOR device are amplitude encoded, also known as on-off keying. One positive side-effect of this is that we are able to implement two output ports, one which acts as an XOR gate and one which acts as its inverse, XNOR.

In order to fully characterize our inverse-designed XOR device, we also designed a device with the same behavior using more conventional simulation design methods. Specifically, we designed a multi-mode interferometer (MMI) device based on a design in Ref. 27. These devices are narrow so they support fewer modes than larger MMI devices and have less complex interference patterns[27]. On one of the inputs, there is a wide region leading up to the device, which is a phase adjuster. This phase adjuster adds $\pi/2$ radians of phase to that input, which adjusts the interference pattern within the device. This phase adjuster is designed to support the fundamental and first guided mode, and could be redesigned to offer a different phase shift[28]. Just as with the inverse-designed device, this conventional device operates with BPSK encoding on the inputs and amplitude encoding on the outputs. One output port acts as an XOR gate, while the other acts as an XNOR gate. The size of the input and output waveguides are the same.

For all of our devices, we designed them to operate at a wavelength of 1310 nm, which is a typical communications wavelength. If we want to target other communications wavelengths, the inverse design can easily be re-run to operate at or near a wavelength of 1550 nm, and adjusting the conventional design to operate at or near 1550 nm can also be done as MMIs are well-understood analytically[29].

In a similar fashion to the XOR optimization, we used Lumerical and LumOpt to generate an inverse design for a y-splitter. In order to save computation time, we only simulated a single arm of the y-splitter and then used a symmetric boundary condition to generate the second arm of the device; connecting to the second output waveguide. Due to the symmetry, we use a target FOM of 0.5 at the single output, meaning half of the power of the original signal should reach the output. This results in a symmetric y-splitter structure that has identical outputs to each output waveguide in the device.

The y-splitter inverse design has one input and two output waveguides with the same cross-section dimensions as the XOR device. The output waveguides were spaced 1.4 μm away from each other. Additionally, the optimization region had a length of 1.8μm and a width of 1.1μm which was located in between the input and top output waveguide, and subsequently mirrored for the lower half of the device, giving an overall device size of 1.8 μm by 2.2 μm.

Like the XOR gate, we first ran a 2D simulation to serve as the starting point for the 3D simulation. In the 2D simulation we specified the initial condition to fill the optimization space with a refractive index halfway between the background and the waveguide indices. By using the 2D results as the initial condition for the 3D optimization, the end results were better than using the other initial conditions.

Unlike the XOR design, there were no major limitations with Lumopt for the y-splitter. We designed a conventional y-splitter using an MMI device as well. The output waveguides were spaced 1.6 μm apart, and the MMI region is 2.8 μm long by 2.2 μm wide.

For the waveguide crossing optimization, we again used a co-optimization method to optimize the crossing in both directions simultaneously. The target FOM in each direction is set to 1. Since both source inputs are the same wavelength, we expect a symmetrical design and identical transmission rate for both output waveguides.

The inverse design waveguides use the same cross-section and wavelength as the other designs. The inverse designed crossing had a much smaller footprint at 2.2 μm by 2.2 μm, while the conventional design had a 3.4 μm by 3.4 μm footprint.

For the waveguide crossing, we only ran the 3D optimization by filling the optimization region halfway between the background and waveguide indices, skipping the initial 2D optimization. We left out the 2D optimization step because the waveguide crossing optimization ran quickly in 3D and gave satisfactory results. Running the 2D optimization first followed by the 3D optimization was overall slower, and the results were similar to optimizing in 3D only. There were no major limitations with Lumopt when designing the waveguide crossing.

The conventional design for the waveguide crossing was also accomplished using overlapping MMI devices in each direction. Each individual MMI region is 3.4 μm by 1.5 μm. By using a multi-mode region, we can reduce the loss and the crosstalk in the region where the waveguides overlap, resulting in better performance than a waveguide crossing of two single-mode waveguides.

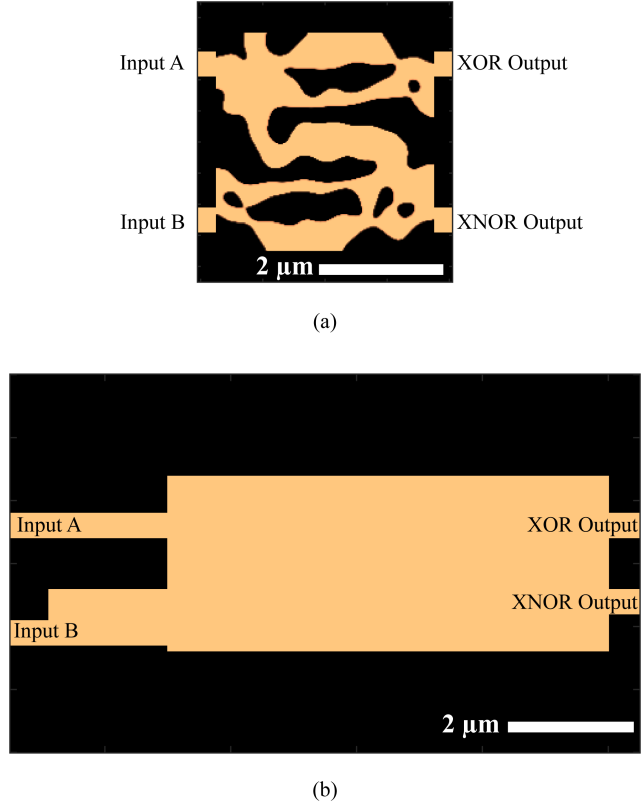

**Fig. 4** Topology of XOR devices designed (a) using an inverse-design tool and (b) using conventional methods. The copper color represents the location of SiO$_2$. Both diagrams are set to the same scale.

## 4 Results

The topology of the design generated from Lumerical's inverse design tool is shown in Fig. 4(a). The generated design is non-intuitive, but in certain respects is similar to a directional coupler. The power inside the device for two different combinations of phase at the input is shown in Fig. 5 (a) and (b), and the **H$_y$** field is shown in Fig. 5 (c) and (d), from which you can see the phase relations between the inputs. From these figures, it is plain to see that the inverse-designed device achieves the desired XOR behavior. When the inputs are in phase with one another, the light is directed to the output port on the bottom of the device. Conversely, when the inputs are π radians out of phase with one another, the top output port is selected. These relations translate to an XOR function with phase-encoding on the inputs and amplitude-encoding on the top output port, and an XNOR function on the bottom output port.

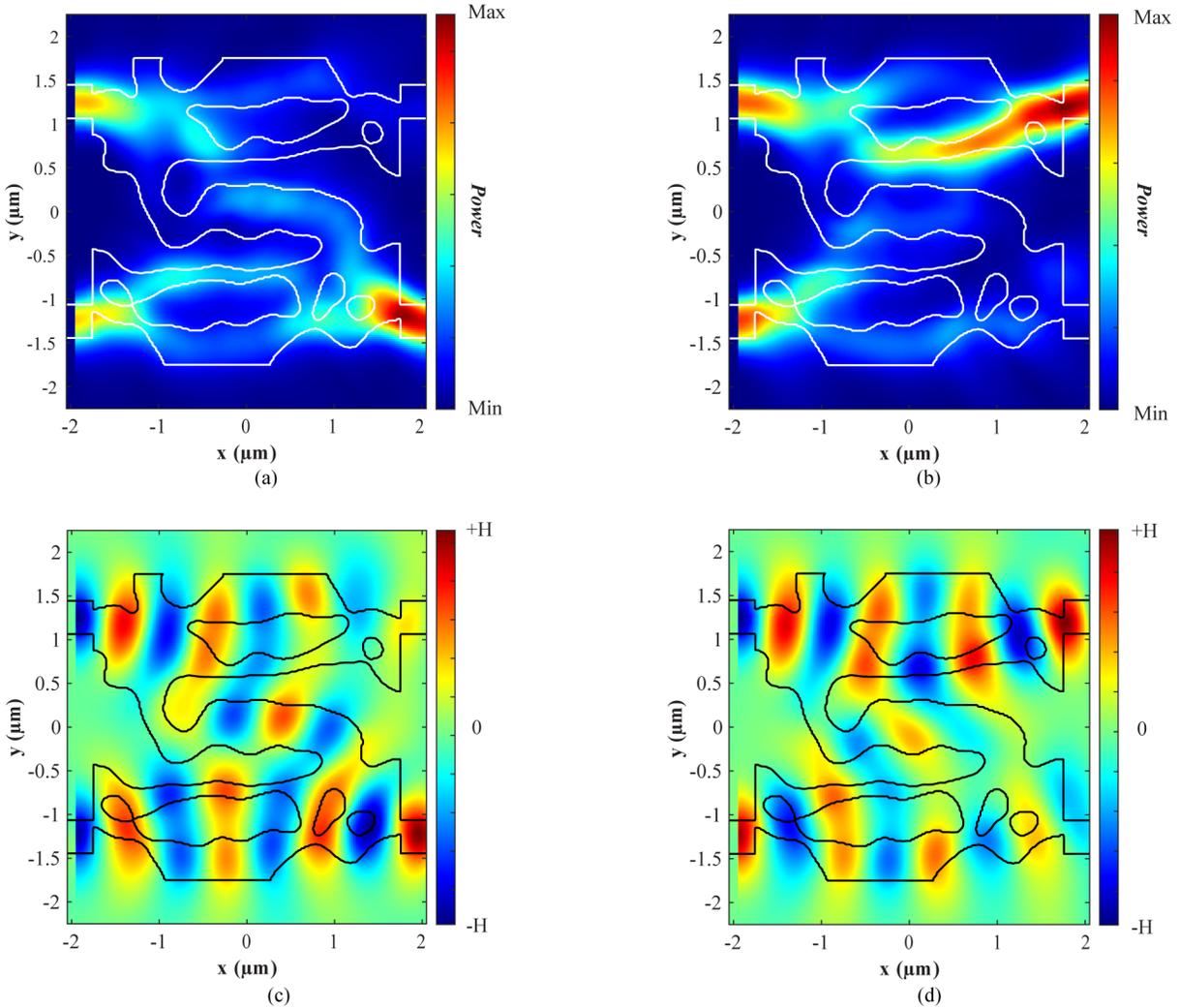

**Fig. 5** Simulation results for the inverse-designed XOR device. The top row shows the power in the device when (a) the inputs are in-phase and (b) the inputs are π radians out of phase. The bottom row shows the $H_y$ field, allowing us to view the phase of the signal, when (c) the inputs are in-phase and (d) the inputs are π radians out of phase. The outline in white on plots (a) and (b) and in black on plots (c) and (d) indicates the topology of the device being simulated.

The topology of the conventionally-designed device is shown in Fig. 4(b). The power inside the device and the $H_y$ fields for different phase combinations at the inputs are shown in Fig. 6 (a) and (b), and Fig. 6 (c) and (d), respectively. Once again, the figures demonstrate the desired XOR behavior. This device behaves in the same way as the inverse-designed device. Inputs with signals that are in phase (00 and 11 cases) cause light to couple into the bottom output port, while inputs with signals π radians out of phase with each other (01 and 10 cases) causes light to couple into the top output port.

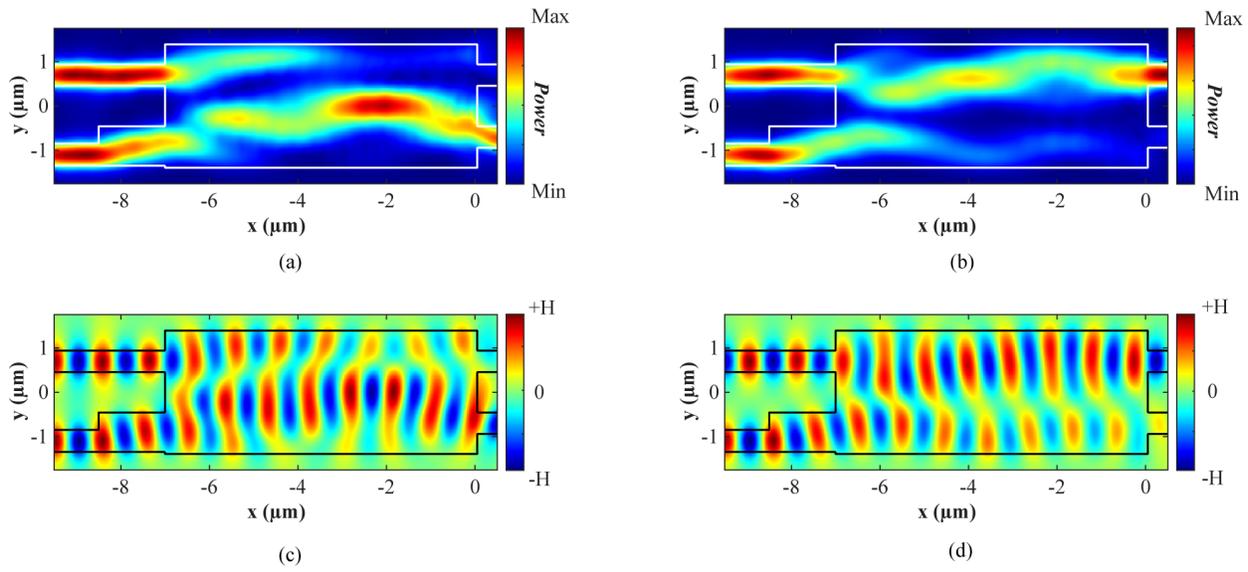

**Fig. 6** Simulation results for the conventionally-designed XOR device. The top row shows the power in the device when (a) the inputs are in-phase and (b) the inputs are π radians out of phase. The bottom row shows the $H_y$ field, allowing us to view the phase of the signal, when (c) the inputs are in-phase and (d) the inputs are π radians out of phase. The outline in white on plots (a) and (b) and in black on plots (c) and (d) indicates the topology of the device being simulated.

In addition to confirming the XOR behavior of the inverse- and conventionally-designed devices, we characterized their performance. Table 2 compares important metrics between both devices. Specifically, we compared insertion loss, extinction ratio, and device area between both devices. The insertion loss is defined as $loss[dB] = -10\log(\frac{P_{out}}{P_{in}})$ where $P_{out}$ is the power coupled into the active output port and $P_{in}$ is the input power to the device. The extinction ratio is defined as $Ratio[dB] = 10\log(\frac{P_{on}}{P_{off}})$ where $P_{on}$ is the power coupled into the output port when it is active and $P_{off}$ is the power coupled into the same output port when it is inactive. Finally, the area of the device was calculated as the active area, excluding the input waveguides. For the inverse-designed devices, this area is the area that was topologically inverse-designed. The inverse-designed region is 3.5 μm by 3.5 μm, which corresponds to an area of 12.25 μm². For the conventionally-designed device, we must include the phase adjuster portion which is essential to achieving the device behavior, so we extended the length to include this component, leading to dimensions of 8.4 μm by 2.8 μm. Even though this device is narrower than the inverse-designed device, the area is 23.52 μm², which is almost twice the area.

Table 2 XOR Gate Performance Comparison

| Device Design | Insertion Loss | Extinction Ratio | Area |
|---|---|---|---|
| Conventional | 8.31 dB (XNOR) / 7.38 dB (XOR) | 8.12 dB (XNOR) / 8.49 dB (XOR) | 23.52 μm² |
| Inverse | 5.24 dB (XNOR) / 5.5 dB (XOR) | 6.34 dB (XNOR) / 7.28 dB (XOR) | 12.25 μm² |

The topology generated for the y-splitters via inverse design and conventional design are shown in Fig. 7 (a) and (b), respectively. The power inside the device for both devices is depicted in Fig. 8 (a) and (b). We compare our designs by calculating the splitting ratio, insertion loss, and area. The splitting ratio is defined as $Ratio(top) = \frac{P_{top}}{P_{top}+P_{bottom}}$ for the top waveguide and $Ratio(bottom) = \frac{P_{bottom}}{P_{top}+P_{bottom}}$ for the bottom waveguide where $P_{top}$ and $P_{bottom}$ are the measured output power in each output waveguide. Given our symmetrical design, both splitting ratios at each output should be the same. The insertion loss is calculated in the same way as the XOR gate, using the total power in both arms as the $P_{out}$.

Table 3 Y-splitter Performance Comparision

| Device Design | Insertion Loss | Splitting Ratio | Area |
|---|---|---|---|
| Conventional | 1.32 dB | 50/50 | 6.16 μm² |
| Inverse | 0.696 dB | 50/50 | 3.96 μm² |

From the figure we can see that the inverse designed y-splitter has more power going to each output waveguide than the conventional design. Table 3 compares important metrics between the designs. We calculate that the inverse design has a substantially lower insertion loss, meaning there is less loss overall in the device than in the conventional design. Additionally, the inverse design is about a third of the length of the conventional design and a little over half of the width of the conventional design, which leads to a smaller area.

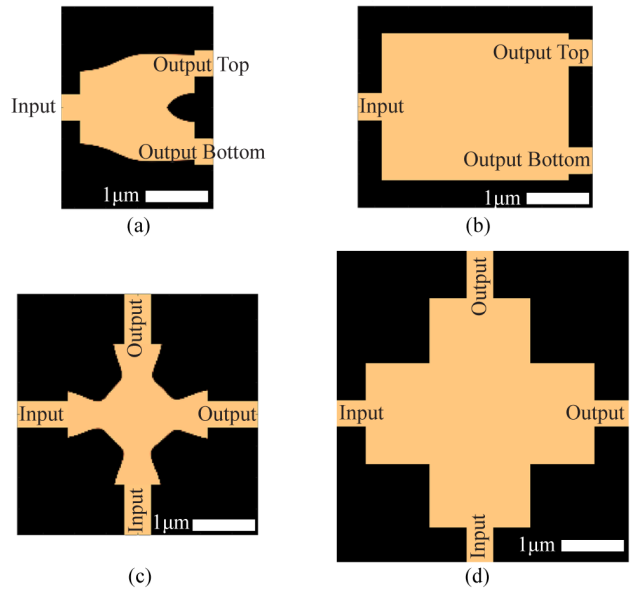

**Fig. 7** Topology of designed devices. (a) Inverse-designed y-splitter, (b) conventionally-designed y-splitter, (c) inverse-designed waveguide crossing, and (d) conventionally-designed waveguide crossing. The copper color represents the location of SiO$_2$. All diagrams are set to the same scale.

The topologies generated for the waveguide crossings via inverse design and conventional design are shown in Fig. 7 (c) and (d), respectively. The power inside the device for both devices is shown in Fig. 8 (c) and (d). We compare the insertion loss, area, and crosstalk of both waveguide crossing designs. The insertion loss for the waveguide crossing is considered for each direction independently, and is calculated as in the other devices. The crosstalk is defined as $Crosstalk[dB] = 10\log\left(\frac{P_{cross}}{P_{through}}\right)$ , where P$_{cross}$ is the output power in the waveguide ports perpendicular to the input and P$_{through}$ is the output power in the desired port.

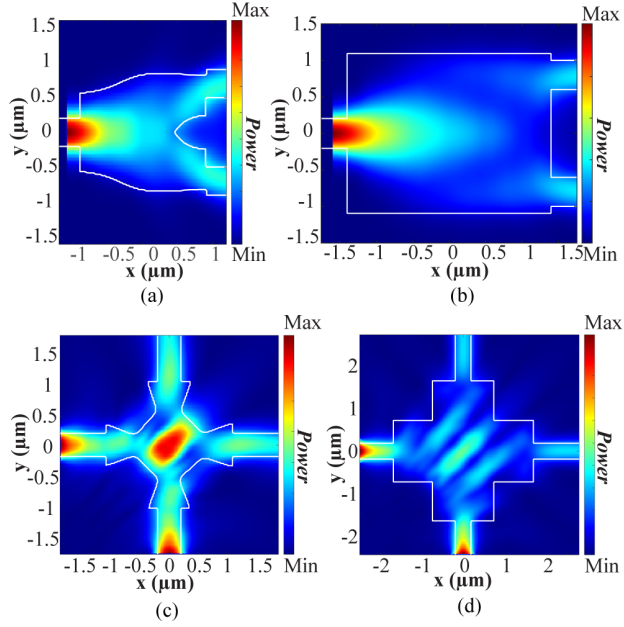

**Fig. 8** Simulation results for the power within various devices. (a) Power inside the inverse-designed y-splitter, (b) power inside the conventionally-designed y-splitter, (c) power inside the inverse-designed waveguide crossing, and (d) power inside the conventionally-designed waveguide crossing. Devices are not to scale. The outlines in white indicate the topology of the device being simulated.

Table 4 compares important metrics between both devices. The inverse designed waveguide crossing has only slightly lower insertion loss but a much lower overall area compared to the conventional design. Additionally, the inverse design has a substantially lower crosstalk.

**Table 4** Waveguide Crossing Performance Comparison

| Device Design | Insertion Loss | Crosstalk | Area |
| --- | --- | --- | --- |
| Conventional | 3.91 dB | -11.8 dB | 11.56 μm$^2$ |
| Inverse | 3.70 dB | -18.4 dB | 4.84 μm$^2$ |

# 5 Discussion

## 5.1 *Evaluating performance of design methods*

For all of our designs the performance of both inverse-designed and conventionally-designed devices is good for a plasmonic device. The insertion loss of these devices is at an acceptable level, and device-specific metrics are also acceptable. For example, in the XOR gate, both devices show a very good extinction ratio, meaning there is little ambiguity in the device behavior. Finally, all designs have a reasonably small overall area for that type of device, a benefit conferred by using plasmonic waveguides.

More interesting features emerge when directly comparing the two design methods for any of the devices. First, the inverse-designed devices generally have a substantially lower insertion loss than the conventionally-designed devices. An exception is the waveguide crossing, where the loss is comparable, but the area is significantly smaller and the crosstalk is substantially reduced for the inverse design. The lower insertion loss is primarily due to the reduced propagation distance in the inverse-designed device. Because propagation losses tend to be high in plasmonic devices, a significant improvement is made by reducing the length of the device.

In the XOR gate device, the lower insertion loss represents a performance tradeoff. In order to achieve a shorter length, the inverse-designed device XOR gate must also have a wider area than the conventionally-designed device. However, this short length probably contributes to the inverse-designed device having a lower extinction ratio than the conventionally-designed XOR gate. While the wider size of the inverse-designed XOR gate may limit how closely these devices could be packed into a single PIC, the overall area of the inverse-designed device is much smaller, so the tradeoff is worth it. Furthermore, even though the extinction ratio is lower, it is still at an acceptable level. The width increase is not a problem for the y-splitter or waveguiding crossing devices. Overall, the inverse-designed devices show improved characteristics over the conventionally designed devices.

*5.2 Limitations of inverse design tools*

For the y-splitter and waveguide crossing designs, there were no significant limitations of the inverse design tool in executing the design. Although the insertion loss did not substantially improve for the inverse-designed waveguide crossing, this is probably more due to the nature of the device itself than a limitation of the tool, and is offset by the substantially lower crosstalk. In contrast, despite the good performance of the inverse-designed XOR gate, there are several drawbacks that arose from limitations in the inverse design tool used. The largest drawback is that we are unable to design the XOR device to have a phase-encoded output using Lumerical's inverse design tools. There is no way to use phase of the light at the output port as a design parameter or FOM. To be clear, designing for phase-encoded outputs is a difficult challenge regardless of the tools used. It is desirable to have phase-encoded outputs to enable arbitrary cascading from outputs to inputs of PSK logic devices. However, if we wish to pursue this approach further, we will either need a second stage to the device that re-encodes the output in phase or need to use a different design approach.

Another drawback arises from the limitation that we can only have one input source. Ideally, we would like to design devices that use higher-order PSK, which uses more phase states to represent encoded multi-bit symbols as opposed to single bits. With one input source, we can represent BPSK inputs by cleverly adapting coupled modes, but this is not possible for higher-order PSK signals. Therefore, to pursue higher-order PSK logic, we will need to use conventional design methods. It is worth noting that higher-order PSK would no longer use binary logic, and therefore would require different kinds of logic gates (for example a MIN and MAX function) or a translation to binary logic with decoder and encoder gates[30]. These are challenging problems worthy of further investigation, and inverse design tools may yet play a role in designing devices for some intermediate stages.

Finally, the Lumerical inverse design tool only works for linear designs. Although this allows us to achieve XOR behavior, if we use phase-encoding we will need to leverage nonlinear optical effects for other logic functions, which are inherently nonlinear. There are several avenues to achieve the desired behavior which are currently under investigation, but none of them involve using Lumerical's inverse design tool. Once again, it may be possible that intermediate stages or sub-components could be designed using these inverse design tools, but it will not be possible to generate a complete device using only inverse design.

# 6  Conclusion

We used Lumerical's inverse design tool with FDTD simulations to design several plasmonic devices, including an XOR gate, y-splitter, and waveguide crossing. The XOR device has phase-encoded inputs and amplitude-encoded outputs, operating with BPSK. Some of the major design decisions were constrained due to limitations of the inverse design tool being used. In order to evaluate the performance of the devices and the benefits of using inverse design, we designed plasmonic devices with comparable operation using conventional simulation methods.

Comparing the performance of the inverse-designed devices to the conventionally-designed devices illustrates that inverse design provides an improvement in insertion loss and device area. For certain devices, such as the waveguide crossing, device area is much more improved than insertion loss. Other device-dependent metrics, such as extinction ratio for the XOR gate, reveal comparable behavior. These performance improvements highlight the usefulness of inverse design methods for creating novel devices, even when their limitations impose additional design constraints. Moving forward, it is likely that the particular limitations of Lumerical's inverse design tool will prevent them from being used by itself for more complex photonic logic design unless improvements and updates are made. However, this and other inverse design tools may still prove useful for intermediate stages and important sub-components, such as y-splitters and waveguide crossings. Inverse design tools may help to reduce the overall area and increase the performance of complex, nonlinear, photonic and plasmonic digital logic devices

*Disclosures*

The authors have no conflicts of interest to disclose.

*Acknowledgments*

The authors thank James Kelly of Chapman University for remote management of simulation tools during the COVID-19 pandemic which enabled this work. This material is based upon work supported by the Air Force Office of Scientific Research under award number FA9550-21-1-0188.

*References*

1. 1.A. Farmani, "Three-dimensional FDTD analysis of a nanostructured plasmonic sensor in the near-infrared range," J. Opt. Soc. Am. B **36**(2), 401 (2019) [doi:10.1364/JOSAB.36.000401].
2. 2.L. Hajshahvaladi, H. Kaatuzian, and M. Danaie, "Design and analysis of a plasmonic demultiplexer based on band-stop filters using double-nanodisk-shaped resonators," Opt Quant Electron **51**(12), 391 (2019) [doi:10.1007/s11082-019-2108-1].
3. 3.L. Cui and L. Yu, "Multifunctional logic gates based on silicon hybrid plasmonic waveguides," Mod. Phys. Lett. B **32**(02), 1850008 (2018) [doi:10.1142/S0217984918500082].
4. 4.Y. Fu, X. Hu, and Q. Gong, "Silicon photonic crystal all-optical logic gates," Physics Letters A **377**(3–4), 329–333 (2013) [doi:10.1016/j.physleta.2012.11.034].

5. 5.M. R. Pav et al., "Ultracompact double tunable two-channel plasmonic filter and 4-channel multi/demultiplexer design based on aperture-coupled plasmonic slot cavity," Optics Communications **437**, 285–289 (2019) [doi:10.1016/j.optcom.2018.12.071].

6. 6.Z. Zeng, P. K. Venuthurumilli, and X. Xu, "Inverse Design of Plasmonic Structures with FDTD," ACS Photonics **8**(5), 1489–1496 (2021) [doi:10.1021/acsphotonics.1c00260].

7. 7.S. Molesky et al., "Inverse design in nanophotonics," Nature Photon **12**(11), 659–670 (2018) [doi:10.1038/s41566-018-0246-9].

8. 8.T. W. Hughes et al., "Adjoint Method and Inverse Design for Nonlinear Nanophotonic Devices," ACS Photonics **5**(12), 4781–4787 (2018) [doi:10.1021/acsphotonics.8b01522].

9. 9.A. Y. Piggott et al., "Inverse design and demonstration of a compact and broadband on-chip wavelength demultiplexer," Nature Photon **9**(6), 374–377 (2015) [doi:10.1038/nphoton.2015.69].

10. 10.P. R. Wiecha et al., "Deep learning in nano-photonics: inverse design and beyond," Photon. Res. (2021) [doi:10.1364/PRJ.415960].

11. 11.D. Melati et al., "Design of Compact and Efficient Silicon Photonic Micro Antennas With Perfectly Vertical Emission," IEEE Journal of Selected Topics in Quantum Electronics **27**(1), 1–10 (2021) [doi:10.1109/JSTQE.2020.3013532].

12. 12."Photonic Inverse Design," Lumerical, <https://www.lumerical.com/solutions/inverse-design/> (accessed 2 July 2020).

13. 13."OptiFDTD Overview," in Optiwave.

14. 14."Photonics Software with FDTD and FETD Engine - Optical Simulations - OmniSim," <https://www.photond.com/products/omnisim.htm> (accessed 13 June 2022).

15. 15."Inverse Design of Photonics | Nanoscale and Quantum Photonics Lab," <https://nqp.stanford.edu/inverse-design-photonics> (accessed 13 June 2022).

16. 16."MATLAB and COMSOL codes for TopOpt-based inverse design in photonics. - TopOpt," https://www.topopt.mek.dtu.dk, <https://www.topopt.mek.dtu.dk/Apps-and-software/Matlab-codes-electromagnetic-topology-optimization> (accessed 13 June 2022).

17. 17.S. Sun et al., "The Case for Hybrid Photonic Plasmonic Interconnects (HyPPIs): Low-Latency Energy-and-Area-Efficient On-Chip Interconnects," IEEE Photonics J. **7**(6), 1–14 (2015) [doi:10.1109/JPHOT.2015.2496357].

18. 18.T. J. Davis, D. E. Gómez, and A. Roberts, "Plasmonic circuits for manipulating optical information," Nanophotonics **6**(3) (2017) [doi:10.1515/nanoph-2016-0131].

19. 19.T. Holmgaard, J. Gosciniak, and S. I. Bozhevolnyi, "Long-range dielectric-loaded surface plasmon-polariton waveguides," Opt. Express **18**(22), 23009 (2010) [doi:10.1364/OE.18.023009].

20. 20.R. Dangel et al., "Polymer waveguides for electro-optical integration in data centers and high-performance computers," Opt. Express **23**(4), 4736 (2015) [doi:10.1364/OE.23.004736].

21. 21.M. Khodami and P. Berini, "Grating couplers for (Bloch) long-range surface plasmons on metal stripe waveguides," J. Opt. Soc. Am. B **36**(7), 1921 (2019) [doi:10.1364/JOSAB.36.001921].

22. 22.W. Bogaerts, M. Fiers, and P. Dumon, "Design Challenges in Silicon Photonics," IEEE J. Select. Topics Quantum Electron. **20**(4), 1–8 (2014) [doi:10.1109/JSTQE.2013.2295882].

23. 23.J. Touch et al., "Digital optical processing of optical communications: towards an Optical Turing Machine," Nanophotonics **6**(3) (2017) [doi:10.1515/nanoph-2016-0145].

24. 24."Photonic Inverse Design Overview - Python API," Lumerical Support, <https://support.lumerical.com/hc/en-us/articles/360049853854-Photonic-Inverse-Design-Overview-Python-API> (accessed 29 September 2021).

25. 25.L. Qian and H. J. Caulfield, "What can we do with a linear optical logic gate?," Information Sciences **176**(22), 3379–3392 (2006) [doi:10.1016/j.ins.2006.02.001].

26. 26.C. M. Lalau-Keraly et al., "Adjoint shape optimization applied to electromagnetic design," Opt. Express **21**(18), 21693 (2013) [doi:10.1364/OE.21.021693].

27. 27.M. Ota et al., "Plasmonic-multimode-interference-based logic circuit with simple phase adjustment," Sci Rep **6**(1), 24546 (2016) [doi:10.1038/srep24546].

28. 28.M. Fukuda et al., "Feasibility of Cascadable Plasmonic Full Adder," IEEE Photonics Journal **11**(4), 1–12 (2019) [doi:10.1109/JPHOT.2019.2932262].

29. 29.L. B. Soldano and E. C. M. Pennings, "Optical multi-mode interference devices based on self-imaging: principles and applications," Journal of Lightwave Technology **13**(4), 615–627 (1995) [doi:10.1109/50.372474].
30. 30.T. Chattopadhyay and J. N. Roy, "All-optical quaternary computing and information processing: a promising path," J Opt **42**(3), 228–238 (2013) [doi:10.1007/s12596-013-0126-0].